\documentclass[11pt,a4paper]{article}

\usepackage[T1]{fontenc}
\usepackage[utf8]{inputenc}
\usepackage[margin=2.5cm]{geometry}
\usepackage{graphicx}
\graphicspath{{figures/}}
\usepackage{booktabs}
\usepackage{array}
\usepackage{amsmath}
\usepackage{setspace}
\usepackage[labelsep=period,font=small]{caption}
\usepackage{url}

\usepackage[super,nomove]{cite}
\makeatletter
\renewcommand{\@biblabel}[1]{#1.}
\makeatother
\newcommand{\refcite}[1]{\cite{#1}}

\begin{document}

\begin{center}
{\Large\bfseries Unified-protocol voxel-level pulmonary embolism annotations for three public CT angiography datasets}

\vspace{1.2em}

Qihang Sun\textsuperscript{1,2,*},
Zhongxiao Liu\textsuperscript{3},
Bailiang Jian\textsuperscript{1,2},
Shenman Qiu\textsuperscript{3},
Jingyuan Wang\textsuperscript{4},
Lei Zhang\textsuperscript{5},
Lixiang Xie\textsuperscript{3},
Jiazhen Pan\textsuperscript{1,2},
Christian Wachinger\textsuperscript{1,2}
\end{center}

\begin{flushleft}
\small
\textsuperscript{1}Laboratory for AI in Medical Imaging, Technical University of Munich, Munich, Germany\\
\textsuperscript{2}Munich Center for Machine Learning, Munich, Germany\\
\textsuperscript{3}Department of Radiology, The Affiliated Hospital of Xuzhou Medical University, Xuzhou, China\\
\textsuperscript{4}Department of Radiology, The Third Affiliated Hospital of Soochow University, Changzhou, China\\
\textsuperscript{5}Department of Radiology, The Affiliated Taizhou People's Hospital of Nanjing Medical University, Taizhou, China\\[0.4em]
\textsuperscript{*}Corresponding author: Qihang Sun (qihang.sun@tum.de)
\end{flushleft}

\vspace{0.8em}

\section*{Abstract}

Reliable clot-volume quantification and subsequent risk assessment in pulmonary
embolism depend on precise segmentation of emboli on computed tomography
pulmonary angiography. Deep learning models for this task must be trained on
accurate voxel-level labels. The three public datasets that provide such labels were
annotated under different protocols, and some of their studies contain
unlabeled emboli or labels that are discontinuous across slices. This Data
Descriptor presents voxel-level pulmonary embolism annotations for 149
of the 166 studies in these datasets. A primary rater drew all annotations under
a single protocol. A thoracic radiologist with more than 20 years of experience
reviewed and revised them. Three raters at three different centers independently
annotated a subset of 15
studies. The subset was selected by source dataset and embolus location.
Technical validation quantifies volumetric agreement with
the source annotations, changes in within-mask attenuation, and inter-rater
agreement on the subset. The dataset is intended to allow segmentation models
to be developed and compared under a common reference standard.

\section*{Background \& Summary}

Acute pulmonary embolism (PE) is a common and life-threatening cardiovascular
disease\refcite{gong,wendelboe,raskob}. Computed tomography pulmonary
angiography (CTPA) is the first-line imaging modality for the rapid assessment
of patients with suspected PE, on which emboli appear as intraluminal filling
defects\refcite{stein,konstantinides}. Because thrombus burden in acute PE is
associated with short-term prognosis, accurate quantification of embolus volume
is important for risk stratification and treatment
planning\refcite{elmenyar,tuzovic,gotta}.

Many deep learning models for PE segmentation have been
reported\refcite{canoespinosa,cheng,pu,trongmetheerat,chen,yang,fan,vadhera2025,nitha,arabian2025,john,zhang,xiao,vadhera2026,arabian2026}.
Several of them were developed only on institutional data that are not publicly
available\refcite{wen,liucamwnet,tang,aydin,dogan,amini,liu2026}.
One reason is that only three public datasets provide voxel-level PE
annotations, CAD-PE\refcite{data_cadpe}, FUMPE\refcite{fumpe} and
READ\refcite{read}, and each was annotated under its own protocol.
Subsequent users of these datasets have described limitations of their
annotations. Fan et al.\ noted that 12 embolus labels are absent from the
released CAD-PE data\refcite{fan}. Chen et al.\ reported substantial
inaccuracies and inconsistencies in the original FUMPE annotations and
re-annotated that dataset\refcite{chen}, and Guo et al.\ identified swapped
slices and axially flipped labels in part of FUMPE\refcite{guo}. Zhang et al.\ described how annotation policy differs across the three datasets.
In FUMPE, subsegmental emboli are omitted. In CAD-PE, partial-volume artifacts in
small vessels are labeled as distal emboli\refcite{zhang}. Such differences are difficult to avoid. Emboli vary widely in shape and are
often small and scattered along the pulmonary arterial tree. Their
three-dimensional delineation also takes far longer than a two-dimensional one.
These factors make systematic differences between annotation teams likely.

Here, the three public voxel-level PE segmentation datasets are combined into a
single collection, FairPE, and re-annotated under one written protocol. Three
trained raters at three different centers independently annotated a
subset of 15 studies. The subset was selected by the most proximal embolus
location and the source dataset. These repeated delineations provide a
multi-rater validation. The collection is
intended to support the development of PE segmentation models and their
comparison under a common reference standard.

\section*{Methods}

\subsection*{Input data}

Three public datasets provide voxel-level annotations of pulmonary embolism on
computed tomography pulmonary angiography (CTPA) and were used as input. CAD-PE
(91 studies, Spain) is distributed through IEEE DataPort\refcite{data_cadpe} and
described in a challenge report\refcite{cadpe_paper}. Its reference standard
was built from regions of interest that radiologists placed on every axial
slice of each visible embolus, expanded into segmentations by an attenuation
threshold with morphological closing and connected-component analysis and then
edited manually; for the 40 studies of the original challenge the segmentations
of three radiologists were merged by simultaneous truth and performance level
estimation (STAPLE), and each of the remaining studies was annotated by at
least one radiologist. FUMPE (35 studies, Iran) is distributed through
figshare\refcite{fumpe}; its reference standard was delineated by one
radiologist and revised by a department head. READ (40 studies, Brazil) is
distributed through figshare\refcite{read}; its reference standard was
delineated by a resident and revised twice by senior experts. All three are released in fully
de-identified form: READ under CC0 1.0, FUMPE under the Creative Commons
Attribution 4.0 International (CC BY 4.0) license, and CAD-PE under the terms
of IEEE DataPort. All three allow redistribution of derived versions.
Table 1
summarizes the demographics, acquisition parameters and reference standards of
the three sources as reported by their original publications.

\subsection*{Case screening and inclusion}

All 166 studies in the three sources were screened before annotation. A study
was excluded if CTPA showed no pulmonary embolism (n = 5), if the reconstructed
slice interval was 3 mm or greater (n = 10), or if imaging artifacts made the
study unannotatable by consensus of all raters (n = 2). The 149 remaining
studies (CAD-PE 76, FUMPE 33, READ 40) were carried forward. The identifiers of
the excluded studies are listed in Fig. 1a, so that the selection can be
reproduced exactly from the public sources.

\subsection*{Annotation protocol}

A written protocol was fixed before annotation began and was applied unchanged
to all 149 studies and by all raters, and is reproduced here in full.

\textit{Reading conditions.} The modality is CTPA in the pulmonary arterial
phase. The primary reading window is the pulmonary artery window (window width
700 Hounsfield units (HU), window level 100 HU), adjusted case by case where
needed.

\textit{Identification criteria.} Pulmonary embolism is identified as an
intraluminal filling defect within the pulmonary arteries, comprising complete
intraluminal hypodensity, partial central or eccentric filling defects (the
railway-track or polo-mint sign), and mural hypodense lesions. All intraluminal
filling defects meeting these criteria are annotated.

\textit{Label definition.} The foreground (label = 1) comprises all filling
defects within the pulmonary arterial lumen that meet the identification
criteria; the background (label = 0) comprises all remaining voxels. Fully
occlusive thrombi are annotated, and where such a thrombus lies at the segmental
level or beyond the annotation is limited to the proximal portion up to the next
bifurcation. Partially occlusive thrombi, mural thrombi and saddle emboli are
annotated. Vessel wall, calcification, contrast medium, pulmonary veins,
parenchymal change and chronic organized lesions are not annotated.

\textit{Boundary rules.} The proximal and distal boundaries are the last slice on
which the thrombus hypodensity is visually separable from the surrounding
contrast medium. In the axial plane, boundaries follow the visual interface
between hypodense thrombus and contrast medium; no single global attenuation
threshold is applied. Subsegmental and more distal lesions are annotated only
when identifiable on two or more consecutive slices. Voxels affected by partial
volume averaging at lesion borders are assigned to the background. Raters
cross-reference adjacent slices to verify thrombus continuity and smooth the
resulting boundaries.

\textit{Difficult studies.} Flow-related and mixing artifacts are not annotated.
Studies with pulmonary arterial attenuation below 200 HU and inhomogeneous
opacification are not annotated and are escalated to the senior reviewing
radiologist. Distal emboli that cannot be reliably distinguished because of
respiratory motion are not annotated.

\textit{Tools.} All delineation was performed in 3D Slicer 5.6.2\refcite{slicer}.
Raters delineated emboli directly from the CTPA images; the source annotations
were displayed only as navigation aids.

\subsection*{Primary annotation and expert review}

The annotation released for all 149 studies (Annotation 1) was produced by the
primary rater, who holds a medical master's degree and had three years of CTPA
experience, and was then reviewed and revised study by study by a
senior thoracic radiologist with more than 20 years of experience in thoracic
imaging. The senior radiologist had authority to modify any voxel of
the primary delineation, and the revised mask is the one distributed.

During annotation, the primary rater and the senior thoracic radiologist
recorded how the source annotation differed from the harmonized annotation in
each study. They used six predefined categories. Four categories cover source
foreground that extends into the pulmonary artery wall, bronchi, lung parenchyma
or pulmonary veins. The fifth covers an embolus present in the harmonized
annotation but absent from the source annotation. The sixth covers anatomically
implausible foreground. The categories are not mutually exclusive. Internal
voids within the source foreground were noted separately. Representative examples are
shown in Fig. 2.

\subsection*{Multi-rater subset}

Fifteen studies, five per source dataset, were selected for independent
re-annotation after stratification by the most proximal level of clot
involvement (central, lobar, segmental or subsegmental), so that the subset
spans all four levels. Three additional raters annotated these studies independently. Each rater was
blinded to the delineations of the others and followed the same written
protocol. Annotation 2 was produced by a
radiologist with 10 years of thoracic CTPA experience. Annotations 3 and 4 were
produced by two CT technologists from different centers. Each had three years of
CTPA experience. The three raters worked at three
different centers. Each of these 15 studies therefore carries four independent
delineations in addition to the source annotation.

\subsection*{Image preprocessing}

The source data were first converted to NIfTI. CAD-PE images and reference
masks were read from NRRD, and the reference values, one per embolus, were
merged into a single foreground label. FUMPE images were assembled from
the DICOM series, and the accompanying MATLAB mask was transposed to the image
axis order and reordered from file-name order into the physical slice order of
the series. READ images were assembled from the DICOM series. They are index-aligned with
the released NIfTI reference. Because of this alignment, each image was written
with the origin and direction of the reference header. Its voxel spacing was
taken from the DICOM tags. These tags are more precise than the rounded spacing
stored in the reference file. Harmonized masks exported from 3D Slicer in NRRD were
binarized on the same grid as the paired image.

Every image and its masks were then cropped to the thoracic region. Anatomical
structures were segmented with the \texttt{total} task of TotalSegmentator
v2.11.0 in fast mode\refcite{totalseg}, the lung labels were merged into one
mask, and the bounding box of that mask was expanded by 10 voxels along each
axis, clipped to the volume, and applied identically to the image and to the
masks, with the affine translated accordingly. Intensities were kept in HU and
were not resampled, windowed or normalized. The conversion and cropping code is
available as stated in the Data Availability section.

\subsection*{Derived measurements and embolus mapping}

The pulmonary artery trunk was segmented with the TotalSegmentator
\texttt{heartchambers\_highres} task, and the
overlap between the trunk mask and the harmonized embolus mask was subtracted
before measurement, so that attenuation of the arterial lumen was sampled only
where the lumen was not occluded. Attenuation in HU and its
standard deviation were then measured within the pulmonary artery trunk and
within each embolus mask. Individual emboli were derived by connected-component
analysis of each mask; components smaller than 2 mm\textsuperscript{3} were
discarded as annotation noise, and the remaining components define the per-study
embolus count and the per-lesion volumes. All attenuation and volumetric
quantities were computed at each dataset's native voxel spacing, without
resampling.

Embolus voxels from all studies were mapped into a common reference space
defined by study READ 20GE. The lung lobes were segmented with the TotalSegmentator
\texttt{total} task and dilated as described above, and the pulmonary artery was
segmented with the \texttt{heartchambers\_highres} task. After removal of
components smaller than 2 mm\textsuperscript{3}, each embolus voxel was assigned
to the lobar or pulmonary artery mask in which it lay. Coordinates were
normalized separately for each structure. Within the bounding box of a
structure, they were rescaled to $[0,1]^3$. They were then mapped into the
bounding box of the same structure in the reference study. The mapped voxels
were accumulated for each structure and smoothed with a Gaussian kernel
($\sigma = 6$ mm).

\subsection*{Validation analysis}

Continuous quantities were tested for normality with the Shapiro--Wilk test and
are reported as mean $\pm$ standard deviation (SD) or as median and
interquartile range [IQR] accordingly. Mean pulmonary artery
attenuation was compared across the three sources with the Kruskal--Wallis test.
Volumetric agreement between the source and harmonized annotations was assessed
by Bland--Altman analysis\refcite{blandaltman}. The mean bias is reported with
its 95\% confidence interval. Paired comparisons of per-study quantities between the
two annotation sets used two-sided Wilcoxon signed-rank tests. Agreement of
case-level embolus volume across the four raters on the multi-rater subset was
quantified by the intraclass correlation coefficient ICC(2,1), a two-way
random-effects, single-measures, absolute-agreement model, with 95\% confidence
intervals\refcite{icc}. Spatial agreement between every pair of annotation sets was quantified by four
measures. These were the Dice similarity coefficient (DSC), the average
symmetric surface distance (ASSD), the normalized surface distance (NSD) at a
1 mm tolerance\refcite{nsd} and lesion-level $F_1$. Lesions were matched with
the scheme of Zhang et al.\refcite{zhang}. In the definitions below,
$\partial P$ and $\partial G$ denote the surface point sets of the two masks
being compared, $d(\cdot,\cdot)$ the
minimum Euclidean distance from a point to a surface in millimeters, and TP,
FP and FN the true positive, false positive and false negative voxels:
\begin{equation}
\mathrm{DSC} = \frac{2\,\mathrm{TP}}{2\,\mathrm{TP} + \mathrm{FP} + \mathrm{FN}}.
\end{equation}
\begin{equation}
\mathrm{ASSD} = \frac{1}{2}\left(
\frac{1}{|\partial P|}\sum_{p \in \partial P} d(p, \partial G)
+ \frac{1}{|\partial G|}\sum_{g \in \partial G} d(g, \partial P)
\right).
\end{equation}
\begin{equation}
\mathrm{NSD}(\tau) = \frac{1}{2}\left(
\frac{\left|\{\,p \in \partial P : d(p, \partial G) \le \tau \,\}\right|}{|\partial P|}
+ \frac{\left|\{\,g \in \partial G : d(g, \partial P) \le \tau \,\}\right|}{|\partial G|}
\right).
\end{equation}
Embolus-level agreement was evaluated with a two-stage volume-aware framework.
In the first stage, each embolus of the first mask was classified as
$\mathrm{TP}_{L}$ if the ratio of its overlap with the second mask to its own
volume ($|P \cap G| / |P|$, where $P$ and $G$ denote the voxel sets of the two
corresponding emboli) met or exceeded threshold $X$ ($X = 1$~pixel, 10\% or
20\%), and as $\mathrm{FP}_{L}$ otherwise. In the second stage, each embolus of
the second mask was considered detected if the ratio of its overlap with
$\mathrm{TP}_{L}$ components to its own volume ($|P \cap G| / |G|$) met or
exceeded the same threshold, and as $\mathrm{FN}_{L}$ otherwise. Lesion-level
precision and recall follow, where $N_{\mathrm{gt}}$ is the total number of
emboli in the second mask:
\begin{equation}
\text{Lesion Precision}(X) =
  \frac{\mathrm{TP}_{L}}{\mathrm{TP}_{L} + \mathrm{FP}_{L}},
\qquad
\text{Lesion Recall}(X) = \frac{\mathrm{TP}_{L}}{N_{\mathrm{gt}}}.
\end{equation}
The lesion-level $F_1$ score at threshold $X$ is the harmonic mean of the two
directional recalls:
\begin{equation}
\text{Lesion F1}(X) =
\frac{2 \cdot \text{Lesion Recall}(X)_{A \rightarrow B}
       \cdot \text{Lesion Recall}(X)_{B \rightarrow A}}
     {\text{Lesion Recall}(X)_{A \rightarrow B}
      + \text{Lesion Recall}(X)_{B \rightarrow A}}.
\end{equation}
When $B$ corresponds to a reference standard, $\text{Lesion
Recall}(X)_{A \rightarrow B}$ and $\text{Lesion Recall}(X)_{B \rightarrow A}$
reduce to lesion precision and lesion recall and the conventional $F_1$ is
recovered; the symmetric formulation allows the same metric to be applied to
two raters of equal status. A STAPLE consensus\refcite{staple} of Annotations
2--4 was computed for visual comparison with Annotation 1, and a STAPLE
consensus of all four annotations was used to order the 15 studies by total
embolus volume for the size-stratified analysis.

\subsection*{Ethics statement}

This work is a secondary analysis of three publicly available, fully
de-identified datasets. Institutional review board approval and informed consent
were therefore not required, and none were obtained. The work was carried out in
accordance with the Declaration of Helsinki and in compliance with the laws,
regulatory frameworks and institutional guidelines applying at the sites where
the annotation was performed. The ethics approvals covering the original
acquisition of the images are reported in the publications of the three source
datasets\refcite{cadpe_paper,fumpe,read}.

\section*{Data Records}

The collection is deposited at Zenodo under the name
FairPE\refcite{data_fairpe}. It will be released under CC BY 4.0 upon acceptance
of this manuscript. This license is compatible
with the terms of the three source datasets. All three are cited alongside this
record and should be cited by anyone who reuses the
data\refcite{data_cadpe,data_fumpe,data_read}. All image and mask files are
gzip-compressed NIfTI-1 volumes. Every mask shares the geometry, voxel spacing
and affine transform of the image volume of the same study, and carries the
values 0 for background and 1 for pulmonary embolism; no other label value
occurs. Volumes are stored at the native spacing of the source dataset and are
not resampled. The volumes are not the full source series. Each image and its
mask were cropped to a box around the lungs, as described in the Methods. The
field of view is therefore smaller than in the source data. Voxel spacing and
Hounsfield values are unchanged, and the affine transform carries the
translation of the crop. File names follow the nnU-Net v2 convention\refcite{nnunet}. An image carries
the modality suffix \verb|_0000|. The corresponding mask carries the same stem
without a suffix. To train with that framework, users place the two directories
under its expected folder names together with a dataset description file.

The record is organized as three top-level directories, one per source dataset.
Each of them holds an \texttt{imgs} directory with the lung-cropped image volumes
and a \texttt{relabeled} directory with the harmonized annotation masks:

\begin{quote}\ttfamily\small
fairpe\_public\_dataset/\\
\hspace*{1em}CADPE/imgs/pe\_001\_001\_0000.nii.gz \ldots\ \ \ (76 files)\\
\hspace*{1em}CADPE/relabeled/pe\_001\_001.nii.gz \ldots\ \ \ \ \ (76 files)\\
\hspace*{1em}FUMPE/imgs/pe\_Patient01\_Patient01\_0000.nii.gz \ldots\ \ (33 files)\\
\hspace*{1em}FUMPE/relabeled/pe\_Patient01\_Patient01.nii.gz \ldots\ \ \ (33 files)\\
\hspace*{1em}READ/imgs/01GE\_0000.nii.gz \ldots\ \ \ \ \ \ \ \ \ \ \ \ \ \ \ (40 files)\\
\hspace*{1em}READ/relabeled/01GE.nii.gz \ldots\ \ \ \ \ \ \ \ \ \ \ \ \ \ (40 files)
\end{quote}

\section*{Data Overview}

Of the 166 screened studies, 149 were included (CAD-PE 76, FUMPE 33, READ 40;
Fig. 1a), and Table 1 lists the demographics, acquisition parameters and source
reference standards of the three datasets. Fig. 1c shows, for each study, the most
proximal level of clot involvement: sub-lobar
studies are least frequent in FUMPE (4/33, 12.1\%) and most frequent in READ
(13/40, 32.5\%). Mapping the harmonized masks into a common reference space
(Fig. 1d) shows a right-lung predominance in CAD-PE and FUMPE and a medial,
lower-zone clustering in all three datasets. Further descriptive statistics can
be derived directly from the released masks.

\section*{Technical Validation}

\subsection*{Comparability of the source images}

Because the three datasets were acquired in three countries on different
scanners, the opacification of the pulmonary arteries was checked before the
annotations were pooled. Mean attenuation of the pulmonary artery trunk showed
no significant difference across the sources (CAD-PE 356.57 $\pm$ 117.15 HU; FUMPE 385.06
$\pm$ 118.84 HU; READ 348.88 $\pm$ 102.77 HU; Kruskal--Wallis H = 1.48,
$P$ = 0.477; Fig. 1b).

\subsection*{Relationship to the source annotations}

Across the 149 studies, Bland--Altman analysis gave a mean difference
(harmonized minus source) of $-5.53$ mL (95\% CI $-7.03$ to $-4.02$ mL; $P<0.001$). Under the
source annotations, per-lesion volume differed markedly between the three
datasets (median 0.18 mL in CAD-PE, 0.67 mL in FUMPE, 0.07 mL in READ), whereas
under the harmonized annotations the three distributions converge (0.13 mL,
0.25 mL and 0.10 mL, respectively; Fig. 3a). Per-study embolus counts
also converge (Fig. 3b): the median count and the interquartile
range decrease in CAD-PE (5 to 4) and READ (8 to 5), while in FUMPE the median
rises from 2 to 3 with a substantially narrower interquartile range (1--5 to
3--4). The two conventions differ in a way that is specific to each source
dataset (Fig. 3c). The main difference is source foreground on the pulmonary
artery in CAD-PE and emboli absent from the source in FUMPE, while READ differs
least.

If a mask contains only intraluminal thrombus, its attenuation should be
comparatively homogeneous; voxels drawn from the vessel wall, from opacified
lumen or from partial-volume boundaries raise its dispersion. Within-mask
attenuation standard deviation was lower under the harmonized annotations in all
three datasets (CAD-PE 131.38 [107.76--164.83] to 65.57 [53.96--85.28] HU; FUMPE
121.46 [81.60--158.82] to 61.03 [52.19--72.28] HU; READ 101.77 [78.72--119.09]
to 76.76 [64.35--95.53] HU; all $P<0.001$; Fig. 3d). Mean within-mask attenuation
was lower in CAD-PE (123.37 [96.31--173.24] to 94.58 [72.24--112.65] HU). It
showed no significant change in FUMPE or READ. The CAD-PE source masks contain
the largest share of opacified high-attenuation voxels. The harmonized
definition excludes these voxels.

\subsection*{Agreement among the four raters}

Across the 15 studies delineated four times, case-level embolus volume showed
excellent agreement between raters, ICC(2,1) = 0.961 (95\% CI 0.926--0.980).
Fig. 4 shows representative studies in the axial and coronal planes. Each view
overlays the source annotation, Annotation 1 and the STAPLE consensus of
Annotations 2--4 on the same slice.

Fig. 5 reports pairwise spatial agreement over the same 15 studies for five
annotation sets. These sets are the four harmonized delineations and the source
annotation. Panels a--f show DSC, ASSD, NSD at 1 mm and lesion-level $F_1$ at
three overlap criteria. Agreement between the source annotation and each of the four harmonized
delineations lies below the range spanned by the pairwise agreements among the
four harmonized delineations, on every metric. Panels g--l repeat the six
measures in two volume groups. The subset was split
at 2 mL by total embolus volume on the STAPLE consensus of all four annotations.
This cut point lies near the center of the volume distribution. It gives a
smaller-volume group (n = 7) and a larger-volume group (n = 8). Pairwise
agreement is higher in the larger-volume group for DSC, ASSD and NSD.
Lesion-level $F_1$ is comparable between the two groups. In the smaller-volume
group, agreement between the source and the
harmonized annotations falls below the pairwise range on every measure.

\section*{Data Availability}

Code for the preprocessing and the evaluation is available at
\url{https://github.com/iblueer777/PEbench}. The dataset will be released upon acceptance of this manuscript.

\clearpage

\begin{table}[htbp]
\centering
\small
\setlength{\tabcolsep}{4pt}
\caption{Patient demographics, acquisition characteristics and source reference
standards of the three public CT pulmonary angiography datasets}
\begin{tabular}{@{}>{\raggedright\arraybackslash}p{3.4cm}>{\raggedright\arraybackslash}p{3.6cm}>{\raggedright\arraybackslash}p{3.6cm}>{\raggedright\arraybackslash}p{4.2cm}@{}}
\toprule
Characteristic & CAD-PE (n = 91) & FUMPE (n = 35) & READ (n = 40)\\
\midrule
Sex, female & 17 (43)\textsuperscript{\dag} & 18 (51) & 27 (68)\\
Sex, male & 23 (57)\textsuperscript{\dag} & 17 (49) & 13 (32)\\
Age, y & 65.0 $\pm$ 18.0\textsuperscript{\dag} & 56.6 $\pm$ 19.7\textsuperscript{\dag} & 55.5 $\pm$ 19.2\\
Country & Spain & Iran & Brazil\\
Inclusion & PE-positive only & PE-positive only (33/35 confirmed) & PE-positive only\\
Scanner & Siemens Somatom Sensation & Mixed & Toshiba Aquilion; GE Revolution\\
Tube voltage, kVp & NR & 120 & 120\\
Tube current, mA & NR & Variable per study & Dose-modulated\\
Pixel spacing, mm & 0.520--0.922 & 0.522--0.785 & 0.572--0.816 (GE 0.600--0.816; Toshiba 0.572--0.782)\\
Slice thickness, mm & NR & 1.0 (33 studies), 2.0 (2 studies) & GE 0.625; Toshiba 1.0\\
Slice interval, mm & 0.5--3.0 & 0.5--1.5 & 0.625--0.8\\
File format & NRRD & DICOM and MATLAB & DICOM and NIfTI\\
Annotation level & Voxel-level & Voxel-level & Voxel-level\\
Source annotation & Semi-automatic from per-slice regions of interest; STAPLE consensus of 3 radiologists for the 40 challenge studies; at least 1 radiologist per study & Manual, 1 radiologist, revised by department head & Manual, resident, double-revised by senior experts\\
\bottomrule
\end{tabular}

\vspace{0.5em}
\begin{flushleft}\footnotesize
Data are mean $\pm$ SD for continuous variables and n (\%) for categorical
variables unless stated otherwise. Counts refer to the full source datasets
before the screening described in the Methods. NR = not reported in the source
publication. PE = pulmonary embolism. STAPLE = simultaneous truth and
performance level estimation. Pixel spacing and slice interval were measured from
the released files. All other demographics and acquisition parameters are as
reported by the source publications.
\textsuperscript{\dag}Computed only from the studies for which the source
publication reports the variable.
\end{flushleft}
\end{table}

\clearpage

\begin{figure}[htbp]
\centering
\includegraphics[width=\textwidth]{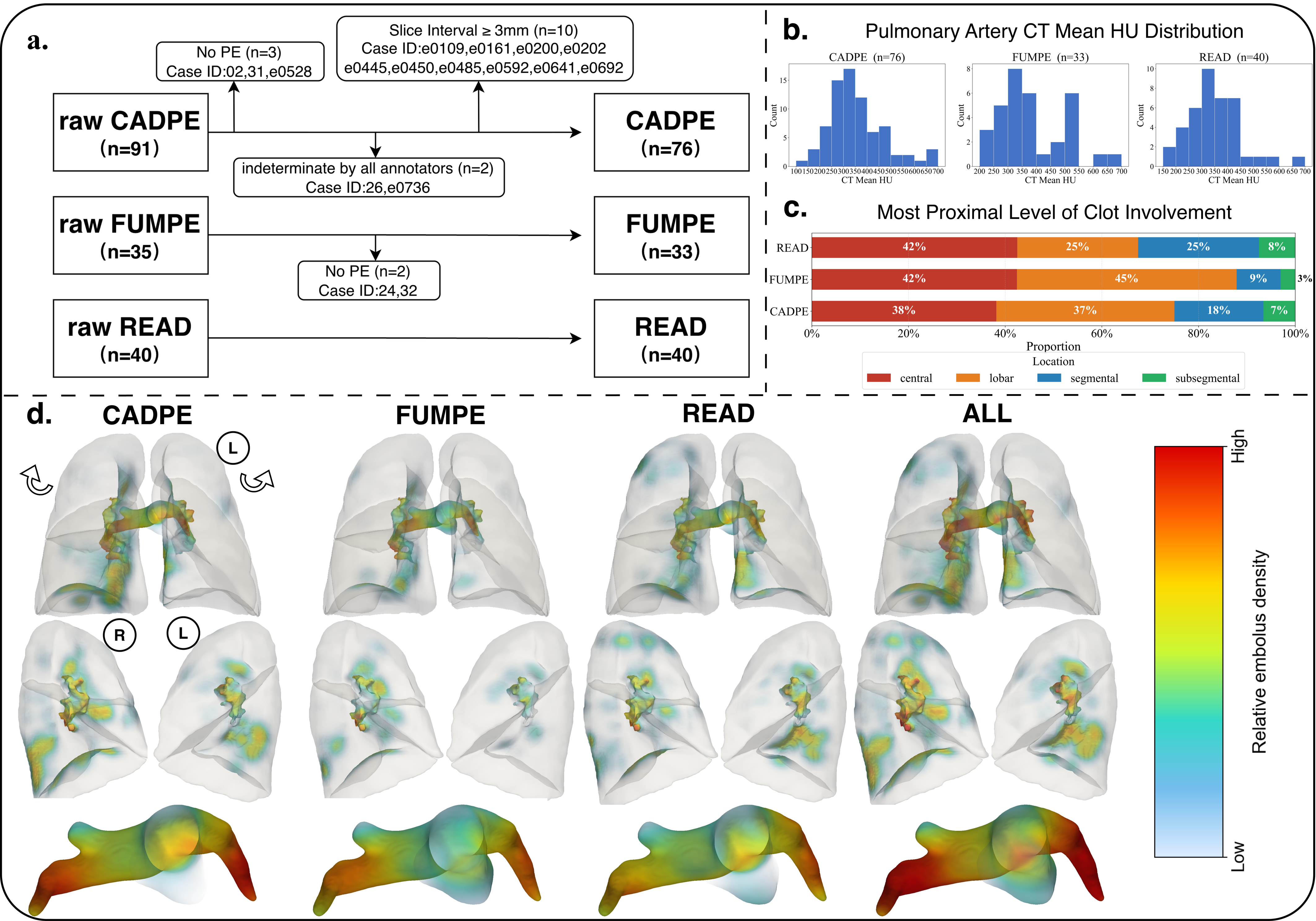}
\caption{Composition and content of the harmonized collection. (a) Screening of
the 166 studies in the three source datasets. Of these, 149 studies were
included (CAD-PE, n = 76; FUMPE, n = 33; READ, n = 40). The identifiers of the
excluded studies are shown. (b) Mean CT attenuation of the pulmonary artery trunk per
dataset, in Hounsfield units. (c) Per-study distribution of the most proximal
level of clot involvement. (d) Embolus density of the harmonized masks
after structure-wise mapping into a common reference space, per dataset and
pooled (ALL): anterior view of the lung volume (top row), medial mediastinal
surface view (middle row) and pulmonary arterial surface (bottom row). Colors
show relative embolus density on a nonlinear scale ($\gamma = 0.25$), with
separate scales for the lung lobes and the pulmonary artery, each shared across
datasets. R and L denote the right and left
lung.}
\end{figure}

\clearpage

\begin{figure}[htbp]
\centering
\includegraphics[width=\textwidth]{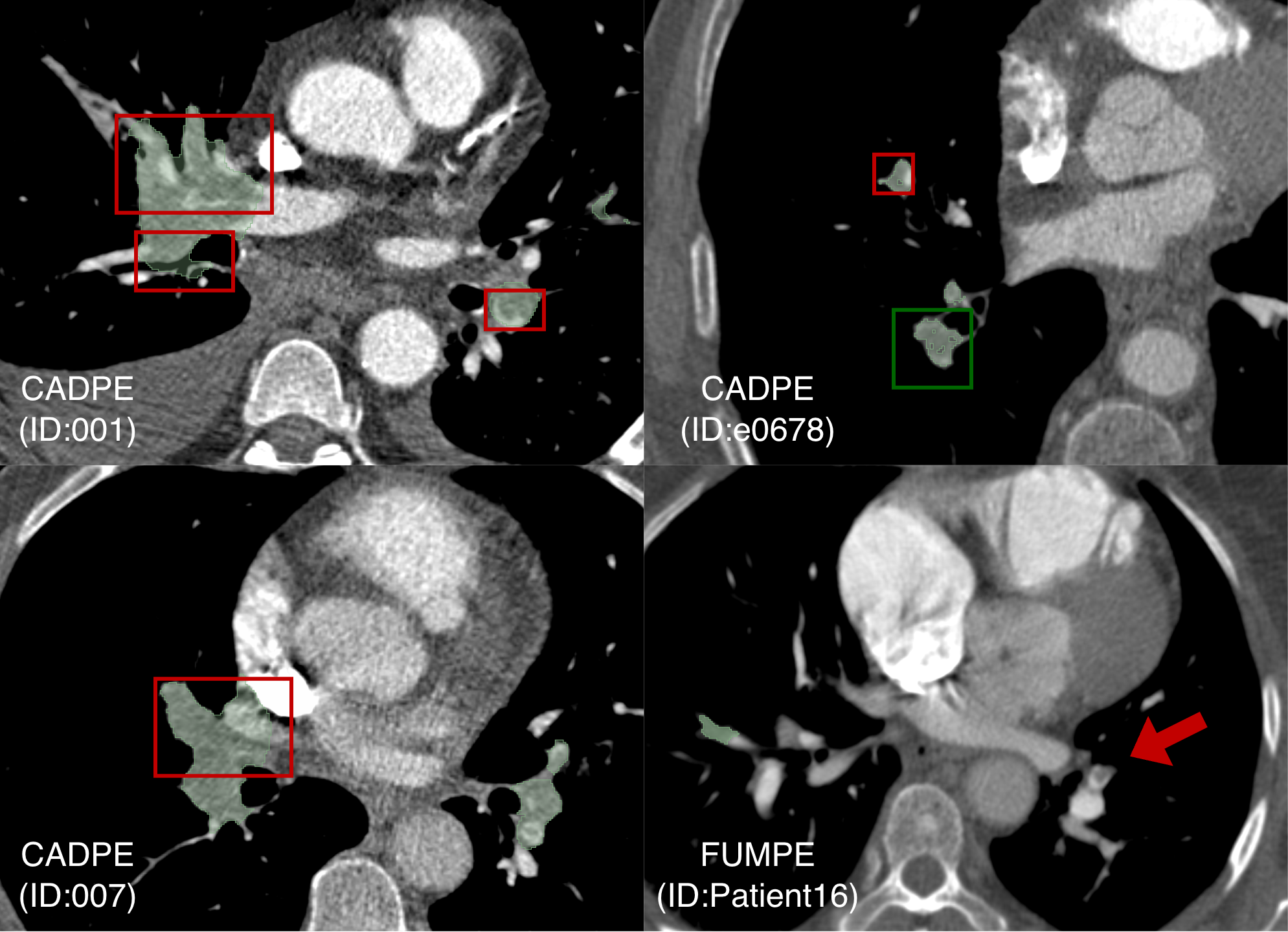}
\caption{Representative differences between the source and the harmonized
annotations. Red boxes mark regions included in the source annotation but
excluded under the harmonized label definition; green boxes mark internal voids
in the source foreground; red arrows mark emboli annotated under the harmonized
definition but absent from the source annotation. Study 001: source foreground
extending into the opacified pulmonary artery. Study e0678: red box, source
foreground extending into the pulmonary artery; green box, void within the
annotated region. Study 007: source foreground extending into a pulmonary vein.
Study Patient16: red arrow, embolus present in the harmonized annotation only.}
\end{figure}

\clearpage

\begin{figure}[htbp]
\centering
\includegraphics[width=\textwidth]{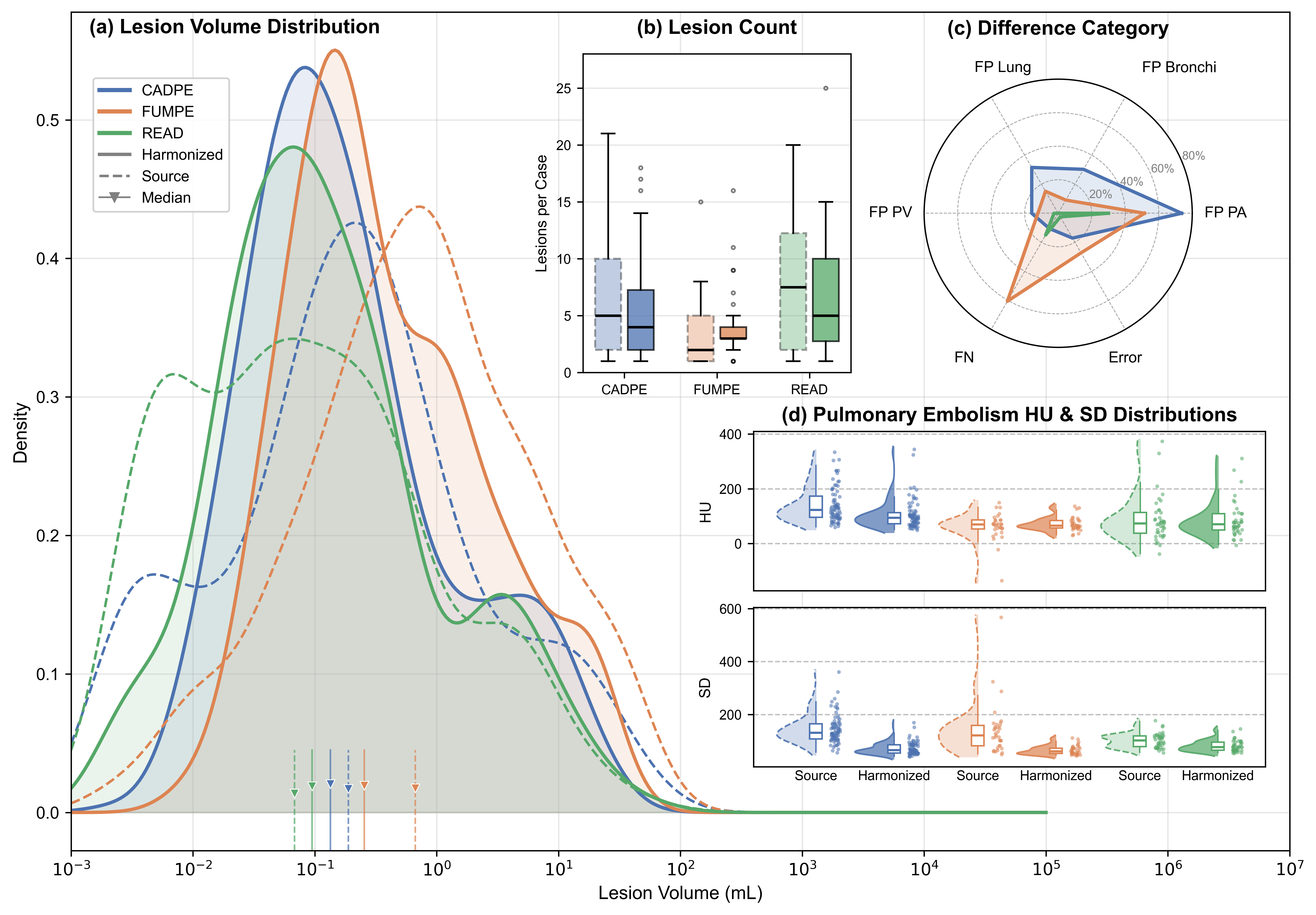}
\caption{Per-lesion volume, per-study embolus count, difference categories and
mask attenuation under the source and the harmonized annotations, by dataset.
(a) Kernel density estimates of per-lesion volume on a logarithmic scale for
the source (dashed) and harmonized (solid) annotations, colored by dataset,
with triangles at the medians. (b) Per-study lesion counts as box plots; light
fill, source; dark fill, harmonized. In all box plots the box spans the
interquartile range, the line marks the median, whiskers extend to 1.5 times
the interquartile range and points beyond are drawn individually. (c) Radar
plot of the predefined difference categories recorded by the two reviewing
raters, as a percentage of studies per dataset. FP PA, FP PV, FP Lung and FP
Bronchi denote source foreground extending onto pulmonary artery, pulmonary
vein, lung parenchyma or bronchi; FN denotes an embolus present in the
harmonized annotation only; Error denotes anatomically implausible foreground.
(d) Per-lesion mask attenuation, mean and standard deviation in Hounsfield
units, under the source and harmonized annotations, shown as combined box,
strip and half-violin plots.}
\end{figure}

\clearpage

\begin{figure}[htbp]
\centering
\includegraphics[width=\textwidth]{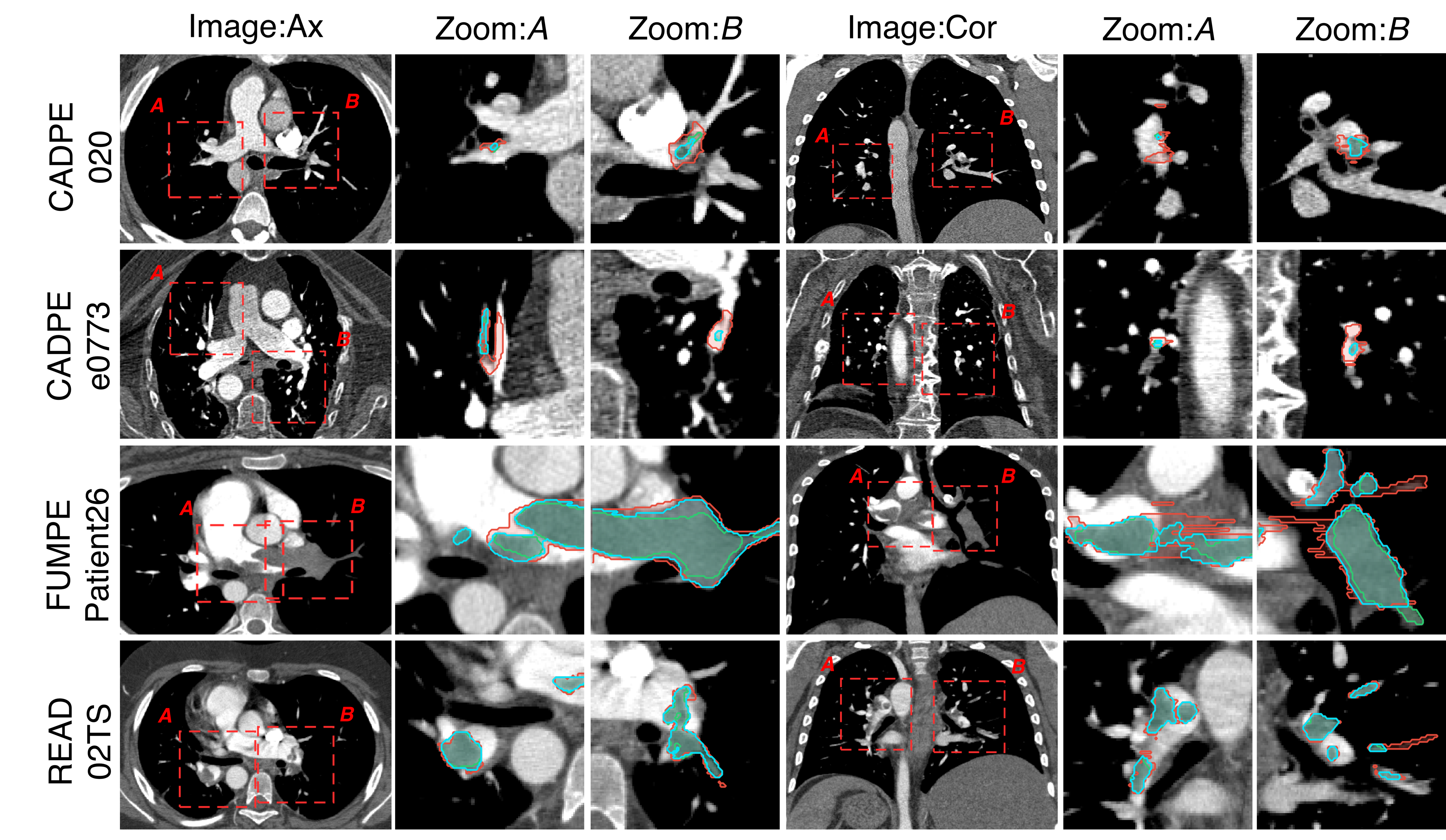}
\caption{Source and harmonized masks with the multi-rater consensus on
representative studies. For each study the same embolus is shown in the axial
and coronal planes; the leftmost image of each plane shows the whole slice and
the two adjacent panels, Zoom A and Zoom B, magnify the two regions outlined by
the red dashed boxes labeled A and B on the full image. Overlays: source
annotation in red, Annotation 1 in green, and the STAPLE consensus of
Annotations 2 to 4 in cyan. Ax denotes axial and Cor denotes coronal.}
\end{figure}

\clearpage

\begin{figure}[htbp]
\centering
\includegraphics[width=\textwidth]{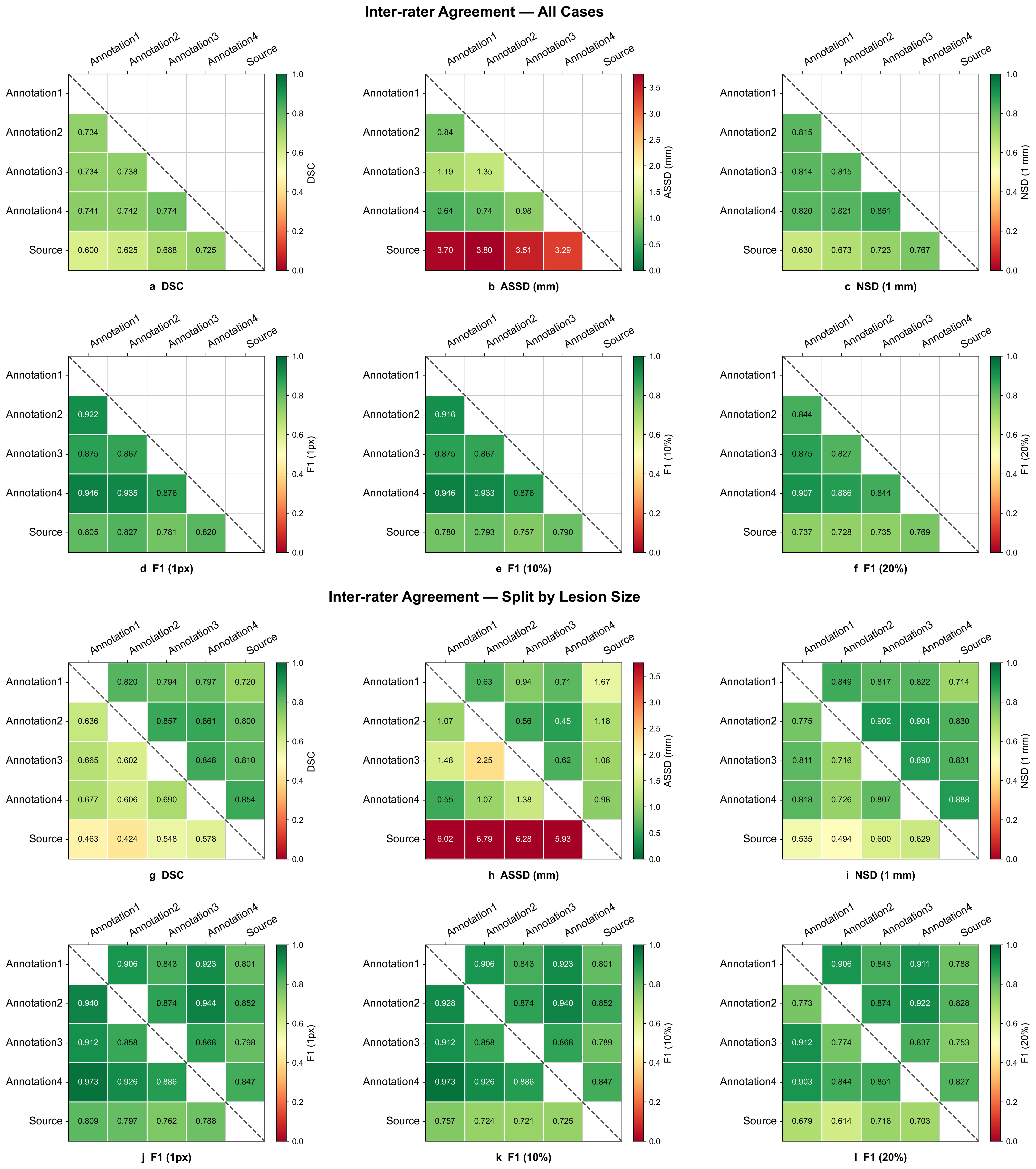}
\caption{Pairwise agreement between the four harmonized delineations and the
source annotation on the 15-study multi-rater subset. Each panel is a
5 $\times$ 5 matrix over Annotations 1 to 4 and the source annotation.
(a--f) Agreement across all 15 studies, as lower-triangular pairwise means:
(a) Dice similarity coefficient, (b) average symmetric surface distance in mm,
(c) normalized surface distance at a 1 mm tolerance, and (d--f) lesion-level
$F_1$ at overlap criteria of 1 pixel, 10\% and 20\%. (g--l) The same six
measures stratified by embolus volume: studies were sorted by total volume on
the STAPLE consensus of all four annotations and split into a smaller-volume
half of 7 studies, shown in
the lower triangle, and a larger-volume half of 8 studies, shown in the upper
triangle, at a cut point of 2 mL: (g) Dice similarity coefficient, (h) average
symmetric surface distance in mm, (i) normalized surface distance at 1 mm, and
(j--l) lesion-level $F_1$ at 1 pixel, 10\% and 20\%. Color bars encode the
scale of each measure; the dashed diagonal in (g--l) separates the two volume
subgroups.}
\end{figure}

\clearpage

\end{document}